\documentclass{JHEP3}
\usepackage[xdvi]{graphicx} 
\usepackage{amsmath}
\usepackage{enumerate}

\title{Plane Waves and Spacelike Infinity}
\author{ Donald Marolf\footnote{E-mail: {\tt marolf@physics.syr.edu}}\ and
Simon F. Ross\footnote{E-mail: {\tt S.F.Ross@durham.ac.uk}}\\
$^*$ Physics Department, Syracuse University, Syracuse, New York
13244 USA\\
$^\dagger$ Centre for Particle Theory, Department of Mathematical
Sciences, University of Durham, South Road, Durham DH1 3LE UK
}

\date{February, 2003}

\abstract{ In an earlier paper, we showed that the causal boundary of
any homogeneous plane wave satisfying the null convergence condition
consists of a single null curve.  In Einstein-Hilbert gravity, this would include any 
homogeneous plane wave satisfying the weak null energy condition.
For conformally flat plane waves such
as the Penrose limit of $AdS_5 \times S^5$, all spacelike curves that
reach infinity also end on this boundary and the completion is
Hausdorff.  However, the more generic case (including, e.g., the
Penrose limits of $AdS_4 \times S^7$ and $AdS_7 \times S^4$) is more
complicated.  In one natural topology, not all spacelike curves have
limit points in the causal completion, indicating the need to
introduce additional points at `spacelike infinity'---the endpoints of
spacelike curves. We classify the distinct ways in which spacelike
curves can approach infinity, finding a {\it two}-dimensional set of
distinct limits.  The dimensionality of the set of points at spacelike
infinity is not, however, fixed from this argument. In an alternative
topology, the causal completion is already compact, but the completion
is non-Hausdorff.  } 
\preprint{SUGP-03/3-2, DCTP-03/11}

\begin{document}
\newtheorem{defn}{Definition}
\newtheorem{thm}{Theorem}

\section{Introduction}

The understanding of the asymptotic structure of spacetimes plays an
important role in many problems in both classical and quantum
gravity. In the past, attention was primarily concentrated on the
formulation of suitable notions of asymptotic flatness and the
exploration of their consequences. There has, however, recently been a
renewal of interest in the careful investigation of the asymptotic
structure of other solutions in the context of the holographic
description of string theory. In particular, the proposed duality
between string theory on a plane wave background and ${\cal N} = 4$
SYM~\cite{BMN} has made it important to understand the structure of these
backgrounds. 

A powerful technique for studying the asymptotic structure is to
construct a suitable completion of the spacetime, adjoining some ideal
points representing the asymptotic behaviour. An elegant method of
constructing such a completion based on the causal structure was
developed in \cite{Geroch,budic,szab1,szab2,MRtop}. This technique was
applied to smooth homogeneous plane waves in \cite{beyond}, where it
was shown that the causal boundary, as defined in
\cite{Geroch,budic,szab1,szab2}, of any homogeneous plane wave
satisfying the null convergence condition\footnote{This is a purely geometric condition which
does not refer to the dynamics of the theory.  In the interesting special case of Einstein-Hilbert gravity, it is equivalent to the weak null energy condition.}
 is a single null curve.  This
generalised a result previously obtained for the special case of the
maximally symmetric (attractive) plane wave in \cite{BN}.

Consideration of the plane wave and other examples has exposed some
defects in the approach to defining the causal completion in terms of
a quotient adopted in~\cite{Geroch,szab1,szab2}. In~\cite{MRtop}, a
new definition of the causal completion $\bar M$ for a general
spacetime $M$ in terms of IP-IF pairs was proposed, and two new
candidate topologies were introduced on this completion\footnote{Also see
\cite{MRtop} for comments on the relation of this scheme to the recent proposal
of \cite{garcia} for constructing Penrose-like diagrams based only on the causal structure.
While this latter construction is far from unique, it may also be of interest in the planewave context.}. Neither of
these topologies is completely satisfactory, but they represent a net
improvement on previous proposals. This new definition was applied to
the homogeneous plane waves in~\cite{MRtop}, and we found that it
reproduces the results previously announced in~\cite{beyond}.

In this paper, we will extend our previous investigations of the
asymptotic structure, applying the topologies defined in~\cite{MRtop}
to investigate curves that approach infinity along spacelike
directions. We will show that taking limits of past and future sets
along spacelike curves generically leads to complicated behavior. In
addition to learning more about the asymptotic structure of the plane
wave spacetime, we hope that the explicit application of the
topologies may help us to better understand the differences between the
two definitions advanced in~\cite{MRtop}.

In one of these topologies, known as $\bar {\cal T}$, most spacelike
curves in these plane waves do not have limit points, even when we
attach the causal boundary to the spacetime. Hence, the causal
completion of the spacetime is non-compact. If we wanted to obtain a
truly compact completion of the spacetime, we would need to adjoin
some additional ideal points reachable only by spacelike curves. Such
points are said to constitute spacelike infinity.

\begin{figure}    
\begin{center}
\includegraphics[width=0.4\textwidth]{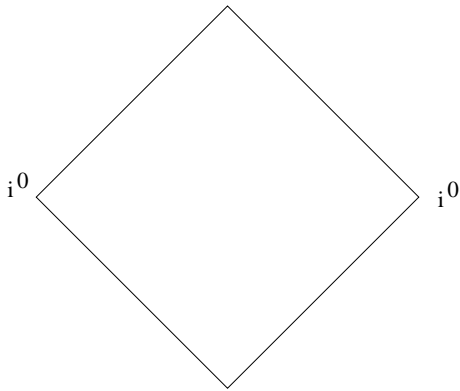}
\caption{A conformal diagram of 1+1 Minkowski space indicating 
spacelike infinity ($i^0$). }\label{i0}
\end{center}
\end{figure}

This is in fact a familiar situation. In the conformal
compactification of Minkowski space, there is a single point $i^0$ in
the boundary which is reachable only by spacelike curves (see
figure~\ref{i0}). If we apply the causal completion technique to
Minkowski space, on the other hand, this point will not be a part of
$\bar{M}$, because no timelike curve reaches it. Hence, the causal
completion is non-compact, and if we want to recover the usual
conformal completion, we have to add in the point $i^0$ `by hand'.

The consideration of spacelike infinity shows that the asymptotic
structure is not the same for all plane waves.  By a detailed study of
general sequences that approach infinity we show below that the
homogeneous plane waves fall into three classes.
\begin{enumerate}
\item The maximally symmetric attractive homogeneous plane wave.  An
example is the BFHP plane wave \cite{BFHP} which arises from a Penrose
limit of $AdS_5 \times S^5$ and has been of much recent interest in
string theory, beginning with the work of BMN \cite{BMN}.

\item Homogeneous plane waves violating the null convergence
condition.  These are unphysical in, e.g., Einstein-Hilbert gravity.

\item All other homogeneous plane waves, including many
\cite{11d,ads4,Cvetic,Bena,Michelson,usc,leo,durham} of interest to
string theory, such as those arising from Penrose limits of $AdS_4
\times S^7$ \cite{ads4}.  Here the causal boundary matches that of
case 1, but using $\bar {\cal T}$ spacelike infinity turns out to be
larger in the sense described below.
\end{enumerate}

The first case is conformally flat and is readily analyzed by
conformal embedding inside a simple globally hyperbolic spacetime (in
practice, the Einstein static universe $S^n \times R$).
In this context, it is clear from the results of
\cite{BN} that all spacelike curves end on the null curve
which forms the causal boundary.  The same observations are readily
seen to be true in the causal completion, using either $\bar {\cal T}$ or
the alternate topology $\bar {\cal T}_{alt}$ also introduced in \cite{MRtop}. 

Case 2 may be discarded.  Thus, the real interest is in case 3.  This
third case contains all smooth homogeneous plane waves that fail to be
conformally flat.  As a result, the Weyl tensor does not vanish for
such spacetimes.  Furthermore, since they are homogeneous, the Weyl
tensor cannot vanish even asymptotically, and these spacetimes cannot
be conformally embedded into a compact region of a smooth
manifold,\footnote{We thank Gary Horowitz for this argument: The
infinite conformal rescaling would require the Weyl tensor of the
spacetime in which we embed to diverge.} so the boundary of such
spacetimes cannot even in principle be addressed by the conformal
method of Penrose \cite{Penrose}. 

In this work, we will show that unlike case 1, the causal completion
$(\bar M, \bar{\cal T})$ in case 3 does not yet contain the limits of
all spacelike curves. That is, the causal completion for these more
general plane waves is non-compact. If we wish to construct a truly
compact completion such as arises from the process of conformal
embedding in case 1, we need to adjoin additional points at spacelike
infinity. We will classify the distinct ways in which spacelike curves
can approach infinity, finding that there is a two-dimensional set of
distinct limits. Unfortunately, this does not directly imply that we
should attach a two-dimensional spacelike boundary. The subtlety is
that distinct limits can arise either from their being different
points at spatial infinity, or from approaching the same point in
different ways. We will not therefore not be able to discuss the
construction of the extended completion in any generality.  It may be
difficult to give a satisfactory definition of such a completion in
the generic case, although we feel that it is clear that one should
exist in sufficiently restrictive circumstances; however, this may
require the use of more information than just the causal information
that our construction of $\bar M$ is based on. 

The discussion above holds for the primary topology $\bar {\cal T}$
introduced in \cite{MRtop}; but \cite{MRtop} also introduces an
alternate topology $\bar{\cal T}_{alt}$, in which more sequences
converge.  We discuss this topology briefly in section \ref{alt}.  It
yields identical results for case 1, while the causal completion of
case 3 becomes compact.  On the other hand, in this case $\bar M$
ceases to satisfy the Hausdorff ($T_2$) separation axiom. We therefore
regard the use of $\bar {\cal T}$ as somewhat more satisfactory than
that of $\bar {\cal T}_{alt}$, since the failure of compactness has a
fairly simple physical interpretation in terms of spacelike infinity,
while we can associate no obvious physical significance with the
failure of Hausdorffness in $\bar {\cal T}_{alt}$. 

We begin in section \ref{prelim} with a summary of useful results from
\cite{beyond}.  In section \ref{rigor}, we consider a specific
spacelike curve, and show that it has no limit point in the causal
completion $(\bar{M}, \bar {\cal T})$. Thus, $(\bar{M}, \bar {\cal
T})$ is not compact.  In section~\ref{sect:i0}, we study the
asymptotic behavior of arbitrary sequences (focusing on those along
spacelike curves); this data will be used to discuss the construction
of spacelike infinity, and to show that $\bar M$ is compact in the
alternate topology $\bar {\cal T}_{alt}$.  We then discuss going to
infinity along spacelike curves in terms of the causal structure of
the spacetime in section~\ref{spacelike}. Finally, we conclude with
some discussion in section \ref{disc}.

\section{Preliminaries}
\label{prelim}

The homogeneous plane waves 
are those solutions~\cite{Brinkmann,BPR,JH,AP,EK,exact} 
to theories incorporating Einstein-Hilbert gravity
for which the metric takes the form
\begin{equation}
\label{mixedmetric}
ds^2 = - 2 dx^+ dx^- - (\mu_1^2 x_1^2 + \ldots + \mu_j^2 x_k^2 - m_1^2
y_1^2 - \ldots - m_{n-k}^2 y_{n-k}^2) (dx^+)^2 + dx^i dx^i + dy^a dy^a,
\end{equation}
in a global coordinate system $(x^\pm,x^i,y^a)$ where each coordinate
ranges over $(-\infty,+\infty)$.  We order the $x_i$ so that $\mu_1
\geq \mu_2 \geq \ldots \geq \mu_n$. Such spacetimes satisfy the null convergence
condition when $\sum_i \mu^2_i - \sum_a m_a^2$ is non-negative.
Thus, we include the case $k=n$ (with no $y^a$ directions) but require
$k \ge 1$.

It was shown in \cite{beyond,MRtop} that the causal completion of
this spacetime has a null curve of ideal points.  Let us call the
boundary points $\bar P_{x^+}$, where we have labeled these points with the
null coordinate $x^+$ used above.  Here $x^+$ takes values in the
extended real line $[-\infty,+\infty]$, which includes the endpoints
$\pm \infty$ and has the topology of the closed interval $[0,1]$.  To
understand the way that these so-called ideal points are attached, let
us recall from \cite{beyond} that a point $p \in M$ lies to the past
of $\bar P_{x^+_0}$ (in the sense that they can be connected by a timelike
curve) if and only if $x^+(p) < x^+_0$.  Similarly, 
$p \in M$ lies to the future of $\bar P_{x^+_0}$ if and only if $x^+(p) >
x^+_0 + \pi/\mu_1$, where $\mu_1$ is the greatest value of $\mu_i$.
Together with the points $p \in M$, these $\bar P_{x^+}$ constitute the
causal completion $\bar M$.

In our analysis below, we will also make use of the characterization
of the future (and past) light cones of interior points $p \in M$.
This is most simply displayed in terms of the future $I^+(0)$ of the
origin ($x=0,y=0,x^+=0,x^-=0$).  From \cite{beyond}, $x \in I^+(0)$ if
$x^+> 0$ and {\it either} of the following two conditions hold:
\begin{equation}
\label{ineq2}
2x^- > \sum_i \mu_i x_i^2 \cot(\mu_i  x^+) + 
\sum_a m_a y_a^2 \coth(m_a  x^+)
\quad {\rm or} \quad x^+ > \pi/\mu_1.  
\end{equation}

We will find it useful below to have an explicit characterization of
when any point $\hat x \in M$ lies to the past of an arbitrary point
$x$.  This can be obtained from the light cone of the origin by using
the symmetries of the plane wave spacetime to translate the origin to
$\hat x$.  Two of the symmetries are simply translations in $x^+$ and
$x^-$.  However, the symmetries that change the values of $x^i,y^a$ are
more subtle.  From e.g. \cite{MPZ}, a symmetry that moves $x^i=0,
y^a=0$ to $x^i=\hat x^i, y^a = \hat y^a$ takes the form
\begin{eqnarray}
x^i &\rightarrow& x^i + \hat x^i \cos(\mu_i x^+) \cr
y^a &\rightarrow& y^a + \hat y^a \cosh(m_a x^+) \cr
x^- &\rightarrow& x^- - \frac{1}{2} \sum_i \mu_i \hat x^i \sin(\mu_i x^+)
[x^i + 
\hat x^i \cos (\mu_i x^+)]  \cr
&+& \frac{1}{2}  \sum_a m_a \hat y^a \sinh(m_a x^+) [y^a
+ 
\hat y^a \cosh (m_a x^+)]    .
\end{eqnarray}

As a result, we see that $\hat x$ lies to the past of $x$ exactly
when $x^+ > \hat x^+$ and either 

\begin{eqnarray}
\label{ineq3}
2( x^- -\hat x^-)   &+&  \sum_i \mu_i \hat x^i x^i  \sin(\mu_i
[x^+ -\hat x^+]) 
 - \sum_a m_a \hat y^a y^a \sinh(m_a [x^+ -
\hat x^+])   \cr
&>& \sum_i \mu_i (x^i-\cos(\mu_i [x^+-\hat x^+]) \hat x^i )^2 \cot(\mu_i [x^+-\hat x^+]) \cr &+& \sum_a
m_a (y^a -  \cosh (m_a [x^+-\hat x^+]) \hat y^a)^2 \coth(m_a [x^+-\hat x^+]) 
\cr &{\rm or}& \ x^+ - \hat x^+ > \pi/\mu_1.
\end{eqnarray}
Similarly, $\hat x$ lies to the future of $x$ exactly when $x^+ < \hat x^+$
and either
\begin{eqnarray} 
\label{futineq3}
2( x^- -\hat x^-) &+& \sum_i \mu_i
\hat x^i  x^i \sin(\mu_i [x^+ -\hat x^+])
-   \sum_a  m_a \hat y^a y^a \sinh(m_a [x^+ - \hat x^+])  \cr &<& 
\sum_i \mu_i (x^i- \cos (\mu_i  [x^+-\hat x^+]) \hat x^i)^2 \cot(\mu_i  [x^+-\hat x^+]) \cr &+& 
\sum_a m_a (y^a - \cosh (m_a  [x^+-\hat x^+]) \hat y^a)^2 \coth(m_a  [x^+-\hat x^+])
\cr &{\rm or}& \ \hat x^+ - x^+ > \pi/\mu_1.  
\end{eqnarray}
Luckily these rather cumbersome expressions will simplify immediately
to (\ref{ineq2}) and the corresponding past light cone in the limits 
considered below.

\section{Non-compactness of $\bar{M}$}
\label{rigor}

We would like to show that the causal completion $\bar{M}$ of the
plane wave (\ref{mixedmetric}) is non-compact in the topology $\bar
{\cal T}$ introduced in~\cite{MRtop}. To do so, we need only show that
there is some infinite sequence of points in $\bar{M}$ which does not
have a limit point, as compactness implies the existence of a limit
point for every infinite sequence.  In this section we consider the common special case where there are at least two harmonic oscillator directions; i.e., where we have both $\mu_1$ and $\mu_2$.  A similar counter-example can be found when $\mu_1$ represents the only harmonic oscillator direction.  We will not present the details as the argument is similar and the associated calculations will be included in the much more general construction of section \ref{sect:i0}.

The topology $\bar {\cal T}$ was introduced \cite{MRtop} on $\bar{M}$
by assuming that for an arbitrary set $\bar{S} \subset \bar{M}$, certain
sets $L^\pm(\bar{S})$ are closed; these sets represent suitable
closures of the chronological future and past of $\bar{S}$. For the
present application to homogeneous plane wave spacetimes, all we need
to know is that an ideal point $\bar P_{x^+}$ of the plane wave will lie in
$L^\pm(\bar{S})$ if and only if its future (past) is a subset of
$I^\pm(\bar{S})$, and that $\bar{S}$ is always contained in
$L^\pm(\bar{S})$.  This is a direct consequence of theorem 8 of
\cite{MRtop} and the observation that $(\bar M, \bar{\cal T})$ is
causally continuous (see appendix).

Consider the set $\bar{S}$ consisting of the spacelike curve defined
by $x^+ = x^- = 0$, $y_a=0$, and $x_i = 0$ for $i \neq 2$. (Since we
only need to produce one counter-example to disprove compactness, we
have considered a particularly simple case. We will discuss the
behaviour of more general curves in the next section.) There
are discrete subsets of this curve which do not have any point of $M$
as a limit point in the topology of $M$ (for example, $\{ x: x_2 = n,
n \in {\mathbf Z} \}$). Since $M$ is homeomorphic to its image in
$\bar{M}$, if such a subset of $\bar{S}$ has a limit point in
$\bar{M}$, it can only be an ideal point.

To exclude this possibility, let us consider the past and future of
this curve. By (\ref{ineq3}), the past of $\bar{S}$ is
\begin{eqnarray}
I^-(\bar{S}) &=& \{x: x^+<0 \mbox{ and either } x^+ < -\pi/\mu_1 \mbox{
or } \exists z \mbox{ such that }\\ 
&& 2x^- + \sum_{i \neq 2} \mu_i x_i^2 \cot(-\mu_i
x^+) + \sum_a m_a y_a^2 \coth (-m_a x^+) \cr &&< \mu_2 (x_2 - z\cos[\mu_2 x^+] )^2 \cot(\mu_2
x^+) - \mu_2 x_2 z \sin(\mu_2 x^+)  \}.  \nonumber
\end{eqnarray}
In particular, we see that $\{x: x^+ < x^+_0 \}$ is a subset of
$I^-(\bar{S})$ when $x_0^+ \le - \pi/2\mu_2$, as when $x^+ < -\pi/(2
\mu_2)$ we can make the RHS of the inequality arbitrarily positive by
taking $z$ large so that the quadratic term dominates. Similarly, the future of $\bar{S}$ is, from
(\ref{futineq3}),
\begin{eqnarray}
I^+(\bar{S}) &=& \{x: x^+ > 0 \mbox{ and either }x^+ > \pi/\mu_1 \mbox{
or } \exists z \mbox{ such that }\\
&&+ 2x^- - \sum_{i \neq 2} \mu_i x_i^2 \cot(\mu_i
x^+) - \sum_a m_a y_a^2 \coth (m_a x^+) \cr &&>  \mu_2 (x_2 - z\cos[\mu_2 x^+])^2 \cot(\mu_2
x^+) - \mu_2 x_2 z \sin(\mu_2 x^+) \},\nonumber
\end{eqnarray}
so $\{x: x^+ > x^+_0 \}$ is a subset of $I^+(\bar{S})$ for $x_0^+ \ge
\pi/(2 \mu_2)$. Thus, the ideal points which lie in $L^-(\bar{S})$ are
$\bar P_{x^+}$ for $x^+ \leq -\pi/(2 \mu_2)$, while the ideal points which
lie in $L^+(\bar{S})$ are $\bar P_{x^+}$ for $x^+ \geq \pi/(2 \mu_2) -
\pi/\mu_1$. Since we assume $\mu_1 > \mu_2$, there are no ideal points
which lie in $L^+(\bar{S}) \cap L^-(\bar{S})$.\footnote{Note that in
the maximally symmetric case, where $\mu_1 = \mu_2 = \mu$, the ideal
point $\bar P_{-\pi/(2\mu)}$ will lie in this set; this is in fact the
limit point of this spacelike curve in this case.} Hence, $[\bar{M}
\setminus L^+(\bar{S})] \cup [\bar{M} \setminus L^-(\bar{S})]$ is an
open set in the topology of~\cite{MRtop} which contains all the ideal
points. Since $\bar{S} \subset L^+(\bar{S}) \cap L^-(\bar{S})$, the
spacelike curve never enters this open set, so no subset of $\bar S$
can have any of the ideal points as a limit point. Thus, there are
infinite discrete subsets of this $\bar S$ which have no point of
$\bar M$ as a limit point. 

Hence, if the homogeneous plane wave $M$ satisfies the null convergence 
condition and is not conformally flat, $\bar{M}$ is not compact.

\section{Limits of spacelike curves for the plane waves}
\label{sect:i0}

We have seen that there are spacelike curves in a general plane wave
which do not have any point of $\bar M$ as an endpoint in the topology
$\bar {\cal T}$. In this section, we will classify the different ways
in which a spacelike curve can go to infinity in terms of the causal
structure. We will apply this classification to show that $\bar M$
{\it is} compact in the topology $\bar {\cal T}_{alt}$. In the
subsequent section, we will discuss what this information tells us
about the points at spacelike infinity we need to add if we want to
form a compact set $\tilde M$ in the topology $\bar {\cal T}$. 

\subsection{Limiting pasts and futures}

We want to study the asymptotic behaviour of general sequences of
points in terms of the causal structure.  One technical complication
arises from the fact that some spacelike curves never leave a compact
region of the spacetime.  Thus, the past and future sets along a
generic curve will not converge.  Recall, however, that our goal is to
add boundary points at spacelike infinity to construct a compact space
$\tilde M \supset \bar M \supset M$.  This requires every sequence of
points $\{x_n\} \subset M$ to have a limit point in $\tilde M$, but
not that every sequence converge.  Note that we need only construct a
limit point for each sequence which fails to have one in $M$ itself.
It turns out to be simplest to directly consider such sequences
instead of spacelike curves per se.

Since our approach to the construction of the completion $\bar M$ was
based on the causal structure, defining points of $\bar M$ by
identifying the spacetime regions to their past and future, it seems
natural to attempt to classify the limits by associating some past and
future sets with them. For limits where we go to infinity along a
timelike curve, there is a very straightforward identification; the
past of the endpoint is naturally identified with the past of the
whole curve. This is how the ideal points were defined
in~\cite{MRtop}: we defined an ideal point by saying $I^+(\bar P) =
I^+[\gamma]$ for any curve $\gamma$ for which $\bar P$ should provide
a past endpoint, and similarly $I^-(\bar P) = I^-[\gamma']$ for any
$\gamma'$ for which it provides a future endpoint.  The past and
future sets $I^\pm(\bar P)$ associated with points on the causal
boundary are thus constructed directly from knowledge of the causal
relations in the original spacetime $M$. The ideal point is then
defined in terms of its future and past; that is, formally as a pair
$(P,P^*)$ of past- and future-sets $P, P^*$.

Spacelike curves are more subtle; the past and future of an endpoint
is generally not the past or future of a curve. We would nonetheless
like to associate limiting past and future sets with sequences which
go to infinity along spacelike curves; in our approach, this seems a
natural way to label distinct limits. In~\cite{MRtop}, a more general
notion of the limit of a sequence of past and/or future sets was
defined\footnote{The definition given here differs slightly from that of \cite{MRtop}, but
the two are equivalent for strongly causal spacetimes $M$ such as the 
homogeneous plane waves.} (not necessarily timelike or spacelike). We can use this to
define the limiting past or future for some sequence $s = \{x_n \}
\subset M$. We say that the past sets $I^-(x_n) \subset M$ converge to
the open past set $I^-(\lim \ s) \subset M$ if and only if
\begin{enumerate}
\item Given $x \in I^-(\lim \ s)$, there exists an integer
$N$ such that $x \in I^-(x_n)$ for all $n > N$, and
\item
Given $x  \notin Cl[I^-(\lim \ s)]$, there exists
an integer $N$ such that, $x \not\in Cl[I^-(x_n)]$ for all $n > N$,
\end{enumerate}
and similarly for open future sets $I^+(x_n)$ converging to some
future set $I^+(\lim \ s)$.   Here $Cl$ denotes the closure in $M$.

If there is a point $\bar P \in \bar M$ such that $I^\pm (\lim\ s) =
I^\pm(\bar P)$, then theorem 8 of~\cite{MRtop} guarantees that the sequence
$s \subset M$ will have $\bar P$ as a limit point in the topology $\bar
{\cal T}$. It seems reasonable to regard the sequence $s$ as having
sensible limiting behaviour in a causal sense under more relaxed
conditions, however. We will therefore say that a sequence $s \subset
M$ converges in $\tilde M$ when both limits $I^\pm(\lim \ s)$ exist.
Similarly, we will say that a sequence $s$ has a limit point in
$\tilde M$ when it has a convergent subsequence.  This gives us some
information about the topology and point-set structure of $\tilde M$:
we are requiring that $\tilde M$ associate some endpoint with any
sequence such that the sets $I^\pm(\lim \ s)$ exist.

\begin{figure}
\begin{center}
\includegraphics[width=0.4\textwidth]{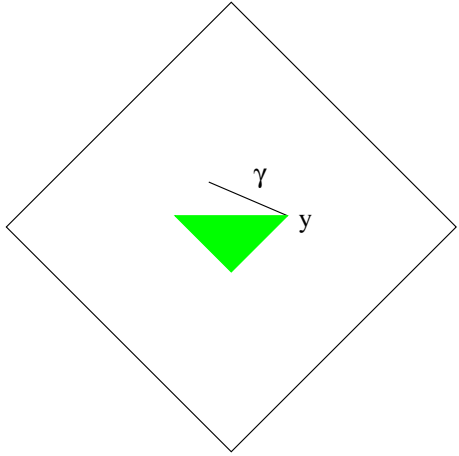}
\caption{In the spacetime constructed by removing the shaded region
  from $1+1$-dimensional Minkowski space, the sets $I^-(x)$ for $x \in
  \gamma$ do not approach $I^-(y)$. }\label{discont}
\end{center}
\end{figure}

We now need to address the physical interpretation of the limiting
past and future sets defined above. Note first that in a general
spacetime, the past or future of the endpoint need not be simply
related to these limiting past and future sets. See figure
\ref{discont} for an example. The assumption that the past (future) of
an endpoint is a limit of pasts (futures) of points in the interior of
a curve is known as causal continuity \cite{HS}. Thus, one can
generally determine the past- and future-sets associated with an
endpoint of a spacelike curve from the data intrinsic to the curve
only if the spacetime is causally continuous. Thus, although we
require that any sequence $s$ such that $I^\pm(\lim \ s)$ exist has an
endpoint in $\tilde M$, we do not want to assume that the $I^\pm(\lim
\ s)$ determine the past and future of this endpoint.

Happily, it is straightforward to show (see appendix) that the
planewave spacetime $\bar{M}$ equipped with the causal boundary of
\cite{beyond} and the topology $\bar {\cal T}$ is causally
continuous.\footnote{Note that $\bar M$ is also weakly distinguishing,
meaning that if $\bar P, \bar Q \in \bar M$ have identical futures
($I^+(\bar P) = I^+(\bar Q)$) and pasts ($I^-(\bar P) = I^-(\bar Q)$)
then they are the same point ($\bar P = \bar Q$).} Thus, we can
conclude that if $I^\pm(\lim \ s)$ are {\it not} the past and future
of some point in $\bar M$, the sequence $s$ will converge to {\it no}
point of $\bar M$. Thus, we need to extend $\bar M$ to construct a
bigger completion $\tilde M$ if there are sequences $s$ such that
$I^\pm(\lim \ s)$ exist, but are not the past and future of some point
in $\bar M$ (however, this will not tell us {\it what} points we need
to add). We will therefore attempt to classify spacelike sequences for
which $I^\pm(\lim \ s)$ exist.

Let us now apply this general approach to the plane waves. Our
approach will be to identify a particular class ${\cal C}$ of
sequences which converge, such that an arbitrary sequence is
guaranteed to contain a subsequence in ${\cal C}$.

It is most transparent to present ${\cal C}$ in several stages.
First, given a sequence of points $\{ x_n \} \subset M$, we would like
to have some control over the behavior of the coordinate sequences
$\{x^+_n \},\{x^-_n \},\{x^i_n \},\{y^a_n \}$.  Thus, we first
restrict to the class ${\cal C}_0$ defined as follows:
\begin{defn}
$s= \{ x_n \} \in {\cal C}_0$ if and only if
each corresponding sequence of coordinates 
$\{x^+_n \},\{x^-_n \},\{x^i_n \},\{y^a_n \}$
converges in the extended real line $[-\infty,+\infty].$
\end{defn}
Since the extended real line has the topology of $[0,1]$ and in
particular is compact, any sequence of points in $M$ contains a
subsequence in ${\cal C}_0$.  Our preferred class of sequences ${\cal
C}$ will be a subclass of ${\cal C}_0$.

We present ${\cal C}$ in terms of three classes, ${\cal C}_M$, ${\cal
C}_{1}$, ${\cal C}_{excep}$ defined as follows:

\begin{defn}
$s= \{ x_n \} \in {\cal C}_M$ if and only if $s$ 
has a limit point in $M$.
\end{defn}
Thus, ${\cal C}_M$ dispenses with the cases already understood.

\begin{defn}
$s= \{ x_n \} \in {\cal C}_1$ if and only if $s \in {\cal C}_0\setminus
{\cal C}_M$ and
each corresponding sequence of ratios 
$\frac{(x^i_n)^2}{x^-_n}$, $\frac{(y^a_n)^2}{x^-_n}$
converges to a finite real number:
$\frac{(x^i_n)^2}{x^-_n} \rightarrow c_i \in {\bf R}$, 
$\frac{(y^a_n)^2}{x^-_n} \rightarrow c_a \in {\bf R}$.
\end{defn}
This is in some sense the generic class of sequences, which will
determine the structure of spacelike infinity.

\begin{defn}
$\{ x_n \} \in {\cal C}_{except}$ if and only if $s \in {\cal C}_0
\setminus {\cal C}_M$,
each corresponding sequence of ratios of pairs of
transverse coordinates
$\frac{x^i_n}{x^j_n}$, $\frac{x^i_n}{y^b_n}$, $\frac{y^a_n}{x^j_n}$,
$\frac{y^a_n}{y^b_n}$,
converges in the extended real line $[-\infty,+\infty]$,
and at least one of the sequences
$\frac{(x^i_n)^2}{x^-_n}$, 
$\frac{(y^a_n)^2}{x^-_n}$ diverges to $\pm \infty$.
\end{defn}
This class contains the `exceptional' sequences which for technical
reasons cannot be analyzed along with those in ${\cal C}_1$.  However,
the study of sequences in ${\cal C}_{except}$ has the same flavor and
yields the same set of limit points.  Again, the compactness of the
extended real line guarantees that any sequence in ${\cal C}_0$ has a
subsequence in ${\cal C}_M$, ${\cal C}_1$, or ${\cal C}_{except}$.

Finally, it is useful to further subdivide ${\cal C}_1$
into two parts ${\cal C}^\pm_1$ based on the behavior of
$x^-_n$:

\begin{defn}
$s= \{ x_n \} \in {\cal C}^+_1$ if and only if
$s \in {\cal C}_1$ and $x_n^- >0$ for all $n$.
\end{defn}
\begin{defn}
$s= \{ x_n \} \in {\cal C}^-_1$ if and only if
$s \in {\cal C}_1$ and $x_n^-  < 0$ for all $n$.
\end{defn}
Since any sequence $\{x_n\} \in {\cal C}_1$ must contain infinitely
many points for which $x^-_n \neq 0$, it contains a subsequence which
lies in ${\cal C}_1^\pm$.  Thus we may take ${\cal C}$ to be ${\cal
C}_M \cup {\cal C}_1^+ \cup {\cal C}_1^- \cup {\cal C}_{except}$.
Note that the long inequalities (\ref{ineq3},\ref{futineq3}) differ
only by the direction of the inequalities.  Though we will consider
the two classes ${\cal C}^\pm_1$ separately, their treatment is
identical as they are simply related by time reversal.

\subsection{Convergence of sequences in ${\cal C}^\pm_1$}

Let us now consider a sequence $s = \{x_n\} \in {\cal C}^+_1$.
We will show that each such sequence is convergent and that
the sets $I^\pm (\lim \ s)$ take the form
\begin{eqnarray}
\label{s-}
I^-(\lim \ s) =  \{x : x^+ < \tilde x^+ - \delta^- \}, \cr
\label{s+}
I^+(\lim \ s) =  \{x : x^+ > \tilde x^+ + \delta^+ \},
\end{eqnarray}
where $x^+_n \rightarrow \tilde x^+ \in [-\infty,+\infty]$.
The parameter $\delta^-$ is the smaller of $\pi/\mu_1$
and the smallest positive solution of
\begin{equation} 
\label{delta+-}
2 =
\sum_i \mu_i c_i \cot(\mu_i  \delta^-) + 
\sum_a m_a c_a \coth(m_a  \delta^-).
\end{equation}
Similarly, $\delta^+$ is the smaller if $\pi/\mu_1$ and
the smallest positive solution of
\begin{equation}
\label{delta++} 
- 2 =
\sum_i \mu_i c_i \cot(\mu_i  \delta^+) + 
\sum_a m_a c_a \coth(m_a  \delta^+).
\end{equation}

To derive this result, notice that since $s \in {\cal C}_0$ but $s
\notin {\cal C}_M$, at least one of the coordinate sequences
$\{x^\pm_n \},\{x^i_n\},\{y^a_n\}$ must approach $\pm \infty$.  If this is
$x^+_n \rightarrow + \infty$, then it is clear that for any $\hat x$ we
have, for large enough $n$, $x^+_n - \hat x^+ > \pi/\mu_1$.  Thus,
$\hat x$ is in the past of all $x^+_n$ with sufficiently large $n$.
Since the plane waves contain no closed causal curves, it also follows
that $\hat x$ is not in the future of such $x_n$.  Thus $s$ converges
and $I^-(\lim \ s) =M, I^+(\lim \ s) = \emptyset$ in agreement with
the statement above.  Similarly, if $x^+_n \rightarrow -\infty$, we
have $I^+(\lim \ s) =M, I^-(\lim \ s) = \emptyset$.

Let us now consider the case $x^+_n \rightarrow \tilde x^+$ with finite
$\tilde x^+$.  Here one of the other coordinate sequences ($x^-_n,x^i_n,y^a_n$) must
diverge.  Since the $c_i,c_a$ are finite and $s \in {\cal C}_1^+$ we
must have $x^-_n \rightarrow +\infty$.  Thus, for large enough $n$,
$\Delta^-_n = x^-_n - \hat{x}^- $ is positive and we may divide the
inequality (\ref{ineq3}) by $\Delta^-_n$ to yield
\begin{eqnarray}
\label{ineq4}
2 &>& 
\sum_i \mu_i \frac{(x^i_n)^2}{x^-_n} \cot(\mu_i  [x^+_n-\hat x^+])
 +  \sum_a m_a \frac{(y^a_n)^2}{x^-_n} \coth(m_a  [x^+_n-\hat x^+]) 
\cr &+& O(x^i_n/x^-_n) + O(y^a_n/x^-_n)
+ O(1/x^-_n) \
{\rm or} \ x^+_n - \hat x^+ > \pi/\mu_1. 
\end{eqnarray}
Note that since some coordinate must diverge, the suppressed
terms are subleading and can be neglected for large $n$.
Taking the limit $n \rightarrow \infty$, we thus arrive at the result
(\ref{s-}) with $\delta^-$ defined by (\ref{delta+-}).

On the other hand, the future condition (\ref{futineq3}) differs from
(\ref{ineq3}) only by the direction of the inequalities while the
definition (\ref{s+}) of $\delta^+$ involves a single extra sign.  As
a result, $I^+(\lim \ s)$ is given by (\ref{s+}) with $\delta^+$
defined by (\ref{delta++}).

This establishes the convergence of $s \in {\cal C}_1^+$.  Sequences
in the class ${\cal C}_1^-$ behave in much the same way, with the only
difference being that $\Delta^-_n$ has the opposite sign.  Thus we find
that any sequence $s \in {\cal C}_1^-$ is convergent and that the
limit again yields sets of the form (\ref{s-}) where $\delta^-$ is the
smaller of $\pi/\mu_1$ and the solution of
\begin{equation} 
\label{delta--}
- 2 =
\sum_i \mu_i |c_i| \cot(\mu_i  \delta^-) + 
\sum_a m_a |c_a| \coth(m_a  \delta^-).
\end{equation}
and $\delta^+$ is the smaller of $\pi/\mu_1$
and the solution of
\begin{equation} 
\label{delta-+}
2 =
\sum_i \mu_i |c_i| \cot(\mu_i  \delta^+) + 
\sum_a m_a |c_a| \coth(m_a  \delta^+).
\end{equation}

\subsection{Limits of ${\cal C}_{except}$}

Let us now consider the class ${\cal C}_{except}$ of exceptional
sequences. As before, if $x^+ \rightarrow \pm \infty$ it is easy to
find the limit, so we focus on the case $x^+ \rightarrow \tilde x^+$
with finite $\tilde x^+$.  Since all ratios of the form
$\frac{x^i_n}{x^j_n}$, $\frac{x^i_n}{y^b_n}$, $\frac{y^a_n}{x^j_n}$,
$\frac{y^a_n}{y^b_n}$, converge in the extended real line
$[-\infty,+\infty]$, we may identify the most rapidly growing
transverse coordinate, which we call $z$.  Note that since at least
one of the ratios $\frac{(x^i_n)^2}{x^-_n}$, $\frac{(y^a_n)^2}{x^-_n}$
diverges to $\pm \infty$, $z^2$ diverges more rapidly than $x^-$.  Let
us therefore divide the inequalities (\ref{ineq3}) and
(\ref{futineq3}) by $z^2$ and take the limit of large $n$ to find that
$\hat x$ lies to the past of $x_n$ for large $n$ 
when $\tilde x^+ > \hat x^+$ and

\begin{eqnarray}
\label{ineq5}
0 &>&
\sum_i \mu_i  \lim_{n \rightarrow \infty}(x^i_n/z_n)^2 \cot(\mu_i  [\tilde x^+-\hat x^+]) + 
\sum_a m_a  \lim_{n \rightarrow \infty}(y^a_n/z_n)^2 \coth(m_a  [\tilde x^+-\hat x^+])
\cr  &{\rm or}& \ \tilde x^+ - \hat x^+ > \pi/\mu_1.  
\end{eqnarray}
Similarly, $\hat x$ lies to the future of $x_n$ for large $n$ exactly when
$\tilde x^+ < \hat x^+$ and
\begin{eqnarray}
\label{futineq5}
 0 &<&
\sum_i \mu_i  \lim_{n \rightarrow \infty} (x^i_n/z_n)^2 \cot(\mu_i  [\tilde x^+-\hat x^+]) + 
\sum_a m_a  \lim_{n \rightarrow \infty} (y^a_n/z_n)^2 \coth(m_a  [\tilde x^+ -\hat x^+])
\cr  &{\rm or}& \ \hat x^+ - \tilde x^+ > \pi/\mu_1.  
\end{eqnarray}

Thus, in this case we again find that any $s \in {\cal C}_{except}$
converges and that $I^\pm(\lim \ s)$ are given by (\ref{s-}),
though this time $\delta^+ = \delta^-$ and this parameter
is the smaller of $\pi/\mu_1$ and the smallest positive solution of
\begin{equation}
\label{delta}
0 =
\sum_i \mu_i  \lim_{n \rightarrow \infty} (x^i_n/z_n)^2 \cot(\mu_i  \delta^\pm) + 
\sum_a m_a  \lim_{n \rightarrow \infty} (y^a_n/z_n)^2 \coth(m_a  \delta^\pm).  
\end{equation}

We have seen that for appropriate $\delta^\pm$ we always
have
\begin{eqnarray}
\label{finallimits}
I^-(\lim \ s) &=& I^-(P_{x_0^+-\delta^-}),  \cr
I^+(\lim \ s) &=& I^+(P_{x_0^+ +\delta ^+ - \pi/\mu_1}).
\end{eqnarray}
Note, however, that in general we will not have $\delta^+ + \delta^- =
\pi/ \mu_1$, so that these sets will not be the past and future of the
same point in $\bar{M}$.  This is seen explicitly in the example
considered in section~\ref{rigor}.

We should note again that in the special case where $\mu_i = \mu_1$
and $k=n$ (so that there are no $m_a$ terms in (\ref{mixedmetric})),
no additional points are needed. This is because the left-hand side of
(\ref{delta++}) changes sign under $\delta^+ \rightarrow \pi/\mu_1 -
\delta^+$.  Since this transformation preserves the range
$[0,\pi/\mu_1]$ of valid $\delta^+$, and since (\ref{delta++}) and
(\ref{delta+-}) agree on the left hand side and differ only by a sign
on the right, this transformation maps $\delta^+$ to $\delta^-$.  For
the (attractive) conformally invariant plane wave, we have identically
$\delta^+ + \delta ^- = \pi/\mu_1$ and in this case the sequence $\{
x_n \}$ converges to the ideal point $\bar P_{x_0^+ -\delta^-}$.  For such
plane waves, the space ${\bar M}$ is already compact without the
addition of further points at spacelike infinity.  This is exactly
what one would expect from the conformal diagram provided in
\cite{BN}.

In the more general case, one might ask whether the ideal point $\bar
P_{x^+_0 + \delta^+}$ associated with the limiting future can ever
{\it precede} the ideal point $\bar P_{x^+_0 - \delta^-}$ associated
with the limiting past.  This would happen only if $\delta^+ +
\delta^- < \pi/\mu_1$.  That this does not occur can be seen by adding
together (\ref{delta++}) and (\ref{delta+-}).  The result is:
\begin{equation}
\label{adddelta}
\sum_i \mu_i |c_i| [\cot(\mu_i  \delta^-)+\cot(\mu_i  \delta^+)] + 
\sum_a m_a |c_a| [\coth(m_a  \delta^-) + \coth(m_a  \delta^+)]= 0.
\end{equation}
Now, the cotangent function is monotonically decreasing on $(0,\pi)$, 
and is antisymmetric about $\pi/2$.  Suppose for a moment
that $\delta^+ \ge \delta^-$.  Then for any term
$ [\cot(\mu_i  \delta^-)+\cot(\mu_i  \delta^+)]$ to be negative or zero,
we must have $\mu_i \delta^+ - \pi/2 \ge \pi/2 - \mu_i \delta^-$.
Thus, $\delta^+ + \delta^- \ge \pi/2 \mu_i
\ge \pi/\mu_1$.  In fact, the boundary value $\delta^+ + \delta^- 
= \pi/\mu_1$ is achieved only when $c_a=0$ and $c_i=0$ for $\mu_i \neq \mu_1$. 
The same is true in the remaining case $\delta^+ \le \delta^-$
due to the symmetry of (\ref{adddelta}) under exchange of $\delta^+$ and $\delta^-$. 
An identical argument can be made for sequences in 
${\cal C}_{except}$.

Finally, note that both $\delta^\pm$ lie in the range $[0,\pi/\mu_1]$.
 Thus, 
$0 < \delta^+ + \delta^- - \pi/\mu_1 \le  \pi/\mu_1$.

\subsection{$\bar M$ is compact in the alternate topology 
$\bar {\cal T}_{alt}$ }
\label{alt}

In addition to the topology $\bar {\cal T}$ addressed thus far, an
alternate topology $\bar {\cal T}_{alt}$ was also introduced in
\cite{MRtop}.  All that we need know about this topology is captured
in theorem 16 of \cite{MRtop}, which states that if $I^+(x_n)
\rightarrow I^+(\bar P)$ for a sequence $\{x_n \} \subset M$ and a
point $\bar P \in \bar M$ with $I^-(\bar P) \neq \emptyset$, then
$\{x_n \}$ converges to $\bar P$.  Similarly, if $I^-(x_n) \rightarrow
I^-(\bar P)$ for a sequence $\{x_n \} \subset M$ and a point $\bar P
\in \bar M$ with $I^+(\bar P) \neq \emptyset$, then again $\{ x_n \}
\rightarrow \bar P$ converges to $\bar P$.

As a result, we can read off the convergence of various classes of
sequences directly from the results of this section.  In fact, together with
the observation that $\delta^+ + \delta^-$ is bounded, equation
\ref{finallimits} tells us that each sequence in the class ${\cal C}$
converges to at least one point in $\bar M$.  Thus, $\bar M$ is
already compact in the topology $\bar {\cal T}_{alt}$.  On the other
hand, we see that most of these sequences converge to two distinct
points $\bar P_{x_0^+-\delta^-}$ and $\bar P_{x_0^+ +\delta ^+ - \pi/\mu_1}$,
since in general $\delta^+ + \delta^- \neq \pi/\mu_1$.  Thus $\bar M,
\bar{\cal T}_{alt}$ is not Hausdorff.

\section{Spacelike infinity}
\label{spacelike}

In the last section, we saw that we could characterise different ways
of going to infinity along spacelike directions in terms of the
limiting past and future sets, using a definition of the limit of a
sequence of past or future sets introduced in~\cite{MRtop}. In the
topology $\bar {\cal T}$, these spacelike sequences have no limit
points in $\bar M$. In defining a larger completion $\tilde M \supset
\bar M \supset M$ by adding points at spacelike infinity by hand, we
need to ask how these limiting past and future sets are related to the
points we add.

First, we should stress that we can never interpret the limiting past
and future sets $I^\pm (\lim s)$ defined in the previous section as
giving the chronological past and future of some point at spacelike infinity. By
definition, there are no timelike curves which approach a point at
spacelike infinity from either future or past. Points at spacelike
infinity are therefore spacelike separated from all points in the
interior of the spacetime, and their pasts and futures in $M$ are
correspondingly empty.

\begin{figure}
\begin{center}    
\includegraphics[width=0.2\textwidth]{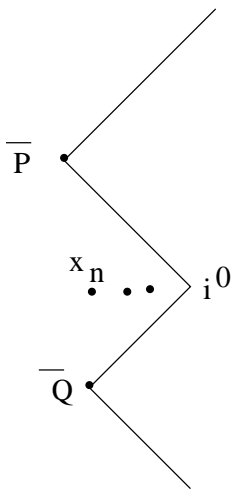}
\caption{Points to the left of the solid line define a subset of 1+1 Minkowski space for which spacelike curves approaching $i^0$ give
nontrivial limiting past and future sets.  }\label{fig:corner}
\end{center}
\end{figure} 

Given that, it may seem surprising that a sequence of points which is
meant to be converging to a point at spacelike infinity can have
non-trivial limiting past or future-sets. However, this is just a
failure of causal continuity in $\tilde M$.\footnote{Thus, we learn
that if we can define a compact $\tilde M$ by adding points at
spacelike infinity, this $\tilde M$ will not be causally
continuous. This does not contradict the result in the appendix, which
only concerns the previously-defined causal completion $\bar M$.}  As
explained previously in section~\ref{sect:i0}, if a spacetime is not
causally continuous, there is no necessary connection between the
limiting past and future sets defined by some convergent sequence and
the past and future sets associated with the endpoint of that
sequence.  A simpler example where this feature is more readily
apparent is shown in figure \ref{fig:corner}.  In this example
$I^+(\lim s) = I^+(\bar P)$, while $I^-(\lim s) = I^-(\bar
Q)$. Nonetheless, the sequence $s = \{ x_n \}$ converges to the point
$i^0$ at spacelike infinity, for which it is clear that $I^+(i^0)=
\emptyset$, $I^-(i^0) = \emptyset$. Somehow in the plane wave this
`corner' phenomenon occurs for a continuum of such pairs $\bar P,\bar
Q$ on the causal boundary.

Thus, we cannot interpret $I^\pm(\lim s)$ as the past and future of
the endpoint we add. However, we would like to ask if we can interpret
them as {\it labeling} the endpoints we need to add.  In general, even
this fails to be true. To see this, consider two simple
$2+1$-dimensional examples constructed from the example in
figure~\ref{fig:corner}. First, let us consider an example where part
of the boundary of our $2+1$ dimensional spacetime is constructed by
rotating the boundary in figure~\ref{fig:corner} about a vertical axis
through $i^0$ (through $\pi/4$, say). This portion of the boundary
then still contains a single point at spacelike infinity
$i^0$. However, we now have a one-dimensional family of distinct
spacelike limits approaching this point, labeled by the limiting past
and future sets $P_\alpha$, $Q^*_\alpha$ which are the images of the
original $P$ and $Q^*$ under rotation through an angle
$\alpha$. Hence, spacelike limits with different $I^\pm(\lim s)$ can
approach the same point.

Next, consider an example where part of the boundary of our $2+1$
dimensional spacetime is constructed by rotating the boundary in
figure~\ref{fig:corner} about a vertical axis through $\bar P, \bar Q$
(through $\pi/4$, say). This now contains a one-dimensional family of
points at spacelike infinity $i^0_\alpha$. However, any spacelike
sequence approaching any one of these points will have $I^-(\lim s)
=P$, $I^+(\lim s) = Q^*$. Hence, spacelike limits with
the same $I^\pm(\lim s)$ can approach different points. 

Finally, a $2+1$ dimensional example with the na\"\i ve one-to-one
relationship between $I^\pm(\lim s)$ and the points at spacelike
infinity can easily be constructed by taking the product of
figure~\ref{fig:corner} with a real line. In this case, we have
one-dimensional families both of $\bar P_z, \bar Q_z$ and $i^0_z$. 

Thus, it seems reasonable to use the limiting past and future sets $I^\pm(\lim \ s)$ associated
with a sequence $s$ to determine if $s$ should become
convergent in $\tilde M$, but perhaps not to determine if different
sequences have the same or different endpoints. Unfortunately, no technology has been
developed to address this latter question within the causal approach.

\section{Discussion}
\label{disc}

The investigation of the asymptotic structure of plane wave spacetimes
is interesting both because of its potential to teach us more about
the holographic relation between string theory and field theory, and
because it presents a challenging case which pushes
the causal boundary technology to its limits. In this paper, we have
shown that the causal completion constructed in~\cite{beyond} is
non-compact in the topology $\bar {\cal T}$ introduced
in~\cite{MRtop}. Thus, to obtain a compact completion, we need to add
additional boundary points at spacelike infinity.

Understanding the limits in spacelike directions seems important, both
because it enables a closer comparison with the more familiar
conformal compactification technique, and because points at spacelike
separation can play an important role in the definition of a duality
between quantum theories. We have therefore investigated the
classification of different spacelike sequences in terms of the causal
structure---how the past and future sets behave as we go to infinity
along the sequence. We used this information to show that in the
alternate topology $\bar {\cal T}_{alt}$ of \cite{MRtop}, the
resulting causal completion $\bar M$ is already compact; however, it
is also non-Hausdorff.

We have attempted to relate the classification of spacelike limits in
terms of the causal structure to some characterization of the points
at spacelike infinity that need to be added to the spacetime. However,
it is not clear to what extent this determines the precise set of points that should be added at spacelike infinity.  Thus, the determination of the appropriate points at spacelike
infinity for the general homogeneous plane waves would require further
work, and will probably involve additional technical choices such as
appeared in the definition of the topologies in~\cite{MRtop}.

It would be interesting to see if one can extend the causal structure
on $\bar{M}$ to a causal structure on the extended compact completion
$\tilde{M}$, once one had a definition of that set.  In \cite{MRtop},
the chronology relation was defined by stating that $\bar P$ is to the
future of $\bar Q$ if and only if $I^+(\bar Q) \cap I^-(\bar P) \neq
\emptyset$.  This definition is naturally extended by saying that
points at spacelike infinity have no chronological relations, as they
are associated with empty past and future sets. Note that this implies
that if $\tilde M$ contains more than one point at spacelike infinity,
it will fail to be weakly distinguishing in this natural chronology.
It is much less clear how one would extend the lightlike (causality)
relation. The example of the conformal compactification of Minkowski
space shows that there may be causality relations which involve points
at spacelike infinity, but it is not clear how these should be
reconstructed.  An answer to this question might also clarify precisely what set of points
should be added at spacelike infinity.

It would be very interesting to relate this discussion to the dual
field theory for string theory on these backgrounds. In particular, it
would be of interest to determine which of the two topologies used
here is most directly associated with properties of the dual gauge
theory.

\acknowledgments 
DM would like to thank James B. Hartle, Gary Horowitz, and Rafael
Sorkin for useful discussions.  DM was supported in part by NSF grant
PHY00-98747 and by funds from Syracuse University. SFR was supported
by an EPSRC Advanced Fellowship.

\appendix

\section{Causal Continuity of $(\bar M, \bar{\cal T})$}
\label{cc}

In this appendix we show that $(\bar M, \bar{\cal T})$ is causally
continuous and, furthermore, that the topology $\bar {\cal T}$ on
$\bar M$ can be characterized by the property that a sequence $\{ x_n
\} \subset \bar M$ converges to $\bar P$ if and only if $\lim I^+(x_n)
= I^+(\bar P)$ and $\lim I^-(x_n) = I^-(\bar P)$.  Note that here we
allow any sequence in $\bar M$, whereas the restriction $\{ x_n \}
\subset M$ was imposed in the main text.  For $x_n \notin M$ we define
$I^\pm(x_n)$ to be $I^\pm_C(x_n) \cap M$, where $I^\pm_C$ is the
chronology defined on $\bar M$ in \cite{MRtop}: $\bar Q \in I^-_C(\bar P)$
if and only if $I^+(\bar Q) \cap I^-(\bar P) \neq \emptyset$.

The concept of causal continuity was introduced in \cite{HS} for
manifolds, and in fact several equivalent definitions were given.
Since $\bar M$ is not a manifold, we have somewhat less structure
here.  Thus, it is no longer clear that the definitions in \cite{HS}
are equivalent.  We take one of these properties, inner and outer
continuity of $I_C^\pm$, to define causal continuity in the present
context.

Recall \cite{HS} that a function $F$ which maps points in $\bar M$ to
subsets of $\bar M$ is called {\it inner continuous} if, for any $\bar
P \in \bar M$ and any compact set $C \subset F(\bar P)$ there is an
open neighborhood $U$ of $\bar P$ such that $C \subset F(u)$ for all
$u \in U$.  Similarly, $F$ is {\it outer continuous} if, for any $\bar
Q \in \bar M$ and any compact set $K \subset \bar M \setminus
Cl{[F(\bar Q)]}$, there is a neighborhood $V$ of $\bar Q$ such that $K
\subset \bar M \setminus Cl[F(v)]$ for all $v \in V$.  Here $Cl$
denotes the topological closure in $\bar M$.  

We now show the following results:

\begin{thm}
$I^-_C$ is inner continuous with respect to $\bar {\cal T}$ for any causal completion $\bar M$ of any strongly causal spacetime $M$.
\label{inner}
\end{thm}
{\it Proof:} Consider $\bar P \in \bar M$ and a compact set $C \subset
I^-_C(\bar P)$.  Let $U = \cup_{I^-(\bar Q) \supset C} I^+_C (\bar
Q)$.  Recall that any $I^-_C(\bar Q)$ is open by Lemma 3 of
\cite{MRtop}.  Thus $U$ is open, and clearly $C \subset I^-_C(u)$ for
all $u \in U$.  Thus we need only show $\bar P \in U$.

Now, $\bar P$
is connected to each point of $C$ by a timelike curve on which there exist
intermediate points $\bar R$. But since $C$ is compact it must be contained
in $I^-_C(R)$ for a finite collection of such points $\bar R_i$. 
Since each such $\bar R_i$ can signal $\bar P$ via a timelike curve in $M$, 
there in fact must be some $r \in M$ with $\bar P \in I^+_C(r)$ that they can all signal
as well.  This $r$ will have $I^-_C(r) \supset C$, so $\bar P \in I^+_C(r)
\subset U$. Thus $I^-_C$ is inner continuous.
$\Box$

\begin{thm}
$I^-_C$ is outer continuous with respect to $\bar {\cal T}$ for the
causal completion $\bar M$ of a homogeneous plane wave satisfying the
null convergence condition.
\label{outer}
\end{thm}
{\it Proof:}  It is useful to first note that, for such $\bar M$, a
point $\bar Q \in L^+(\bar P)$ if any
only if $\bar P \in L^-(\bar Q)$, where $L^\pm$ are defined in \cite{MRtop}.  
Furthermore, these are the closures of the sets $I^+_C(\bar P),I^-_C(\bar Q)$.
We also recall from \cite{MRtop} that $\bar R \in L^-(\bar Q)$ implies
$I^-_C(\bar R) \subset L^-(\bar Q)$.

Now consider any $\bar Q \in \bar M$ and any compact $K \subset \bar M
\setminus L^-(\bar Q)$ and note that the boundary point $\bar P_{-\infty}$
lies in $I^-_C(\bar Q) \subset L^-(\bar Q)$.  As a result, it does not
lie in $K$.  Furthermore, note that there is a past-directed timelike
curve $\gamma_{\bar R}$ through $M$ from any point $\bar R \in K$ to
$\bar P_{-\infty}$.  Since $\bar R \notin L^-(\bar Q)$, we have $I^-(\bar R) \not\subset
L^-(\bar Q)$ from the definition of $L^-$ in \cite{MRtop}.  Thus, some points on $\gamma_{\bar R}$ will lie in $\bar M \setminus
L^-(\bar Q)$ and we may associate each $\bar R \in K$ with some
$\bar R' \in \gamma_{\bar R} \setminus L^-(\bar Q)$.  Now, $K \subset
\cup I^+_C(\bar R')$ so in fact $K \subset \cup_i I^+_C(\bar R'_i)$,
for a finite subcollection $\{ \bar R'_i \}$ since $K$ is compact.

Recall that $L^+(\bar R'_i) \supset I^+_C(\bar R'_i)$ and let $V =
\bar M \setminus [ \cup_i L^+(\bar R'_i)]$.  The set $V$ is open since
the collection $\{ \bar R'_i \}$ is finite.  Since $\bar R'_i \notin
L^-(\bar Q)$, it is clear that $\bar Q \in V$.  Note that any $v \in
V$ must have $L^-(v) \cap K = \emptyset$, else some $\bar R'_i$ would
lie in $L^-(v)$ and $v$ would lie in $L^+(\bar R'_i)$.  Thus $I^-_C$ is outer continuous. $\Box$

\begin{thm}
\label{char}
For homogeneous plane wave spacetimes satisfying the null convergence
condition, a sequence
$\{ x_n \} \subset \bar M$ converges to $\bar P$ in $\bar {\cal T}$
if and only if $\lim I^+_C(x_n) = I^+_C(\bar P)$ and $\lim I^-_C(x_n) = I^-_C(\bar P)$.
\end{thm}
{\it Proof:} Theorem 8 of \cite{MRtop} shows that convergence of the
past and future sets $\lim I^\pm_C(x_n)$ implies convergence in $\bar
{\cal T}$ when $\{ x_n \} \subset M$.  For homogeneous plane waves
satisfying the null convergence condition, the case where boundary points
appear in this sequence is easily handled by inspection.

For the converse, suppose that $\{ x_n \} \subset \bar M$ converges to
$\bar P$ in $\bar {\cal T}$.  Consider any $\bar Q \in I^-(\bar P)$
and note that $\{ \bar Q \}$ is compact.  Then from theorem
\ref{inner}, we see that $\bar Q \in I^-_C(x_n)$ for sufficiently
large $n$.  Similarly, consider any $\bar R \notin Cl[I^-_C(\bar P)]$
and note that $\{ \bar Q \}$ is compact.  Then from theorem
\ref{outer}, we find that $\bar Q \in I^-_C(x_n)$ for sufficiently
large $n$.  $\Box$

Thus, we may characterize the topology $\bar {\cal T}$ by theorem
\ref{char}.

\end{document}